\documentclass[10pt,a4j]{article}
\usepackage[dvips]{graphicx}
\usepackage{mathrsfs}
\usepackage{eucal}
\usepackage{amsmath,amsthm,amssymb}
\usepackage{wrapfig}
\usepackage[dvips]{color}
\usepackage{cite}
\usepackage{framed}
\usepackage{tabularx}
\usepackage[sectionbib]{chapterbib}
\usepackage{threeparttable}
\definecolor{shadecolor}{gray}{0.80}
\DeclareMathVersion{chem}
\SetSymbolFont{letters}{chem}{OT1}{cmr}{m}{n}

\begin{document}

\renewcommand{\figurename}{\small{Fig.}~}
\renewcommand{\thefootnote}{$\dagger$\arabic{footnote}}

\begin{flushright}
\textit{Molecular Weight Dependence of Excluded Volume Effects}
\end{flushright}
\vspace{1mm}

\begin{center}
\setlength{\baselineskip}{25pt}{\LARGE\textbf{Molecular Weight Dependence of Excluded Volume Effects}}
\end{center}

\vspace{-7mm}
\begin{center}
\setlength{\baselineskip}{25pt}{\Large\textbf{From Concentration Dependence of Excluded Volume Effects}}
\end{center}
\vspace{0mm}

\vspace*{0mm}
\begin{center}
\large{Kazumi Suematsu} \vspace*{2mm}\\
\normalsize{\setlength{\baselineskip}{12pt} 
Institute of Mathematical Science\\
Ohkadai 2-31-9, Yokkaichi, Mie 512-1216, JAPAN\\
E-Mail: suematsu@m3.cty-net.ne.jp,  Tel/Fax: +81 (0) 593 26 8052}\\[8mm]
\end{center}

\hrule
\vspace{0mm}
\begin{flushleft}
\textbf{\large Abstract}
\end{flushleft}
Molecular weight dependence of excluded volume effects is examined. The swollen-to-unperturbed coil transition point shifts to lower concentration range with increasing molecular weight (M$_{w}$). It is shown that in the limit of M$_{w}\rightarrow \infty$, the excluded volume effects should vanish in all concentration range, except for the only one point, $C_{0}=0$. Quite in contrast, for short chains, the excluded volume effects never disappear even at the melt state.\\[-3mm]
\begin{flushleft}
\textbf{\textbf{Key Words}}:
\normalsize{Excluded Volume Effects/ Swollen-to-unperturbed Coil Transition/ Molecular Weight Dependence}\\[3mm]
\end{flushleft}
\hrule
\vspace{3mm}
\setlength{\baselineskip}{13pt}
\section{Introduction}
A polymer solution is in itself an inhomogeneous system of segment concentration, which is due to the fact that monomers are joined by chemical bonds. The concentration is highest at the center of gravity, but decreases rapidly with increasing distance from the center. This inhomogeneity gives rise to the gradient of the Gibbs free energy between the inside of a coil and the outside, and this is the origin of the excluded volume effects. Our fundamental idea is thus that the excluded volume effects are manifested as a result of the wild inhomogeneity of polymer solutions. 

On the basis of the above concept, the theory of the excluded volume effects\cite{Flory, Kazumi} leads us, in a natural fashion, to the following equation:
\begin{equation}
\alpha^{5}-\alpha^{3}=N^{2}\frac{V_{2}^{\,2}}{V_{1}}\left(1/2-\chi\right)\left(\frac{\beta}{\pi}\right)^{3}\iiint\left(G_{hill}^{\,2}-G_{valley}^{\,2}\right)dxdydz\label{1-1}
\end{equation}
with $N$ being the number of segments, $V$ the volume (the subscripts 1 and 2 signify solvent and segment, respectively), $\chi$ the enthalpy parameter defined by $\Delta H\propto\chi$, and $\beta=3/2\langle s^{2}\rangle_{0}$ ($\langle s^{2}\rangle_{0}$ denotes the mean square radius of gyration of an unperturbed chain).

In eq. (\ref{1-1}), $G$ is a function associated with segment concentration at the coordinate $(x, y, z)$ (the subscripts $hill$ and $valley$ signifying concentrated and dilute regions, respectively) and has the form
\begin{equation}
G(x,y,z)=\sum_{\{a,b,c\}}\exp\{-\beta[(x-a)^{2}+(y-b)^{2}+(z-c)^{2}]\}\label{1-2}
\end{equation}
Now an additional new term
\begin{equation}
J=\iiint\left(G_{hill}^{\,2}-G_{valley}^{\,2}\right)dxdydz\label{1-3}
\end{equation}
directly related with the concentration fluctuation has been introduced. The term $J$ is a direct manifestation of the wild inhomogeneity of polymer solutions. When $J$ diminishes, the excluded volume effects must also diminish accordingly.
\section{Simulation}
In this report, we solve eq. (\ref{1-1}) as a function of molecular weights, modeling polystyrene solutions in carbon disulfide (PSt in CS$_{2}$). The employed physico-chemical parameters are listed in Table \ref{table1}.
\begin{table}[!htb]
\vspace{-2mm}
\caption{Basic parameters of polystyrene solution\label{table1}}
\begin{center}
\vspace*{-1.5mm}
\begin{tabular}{l l c r}\hline\\[-1.5mm]
& \hspace{10mm}parameters & notations & values \,\,\,\,\\[2mm]
\hline\\[-1.5mm]
polystyrene (PSt) & volume of a solvent (CS$_{2}$) & $V_{1}$ & \hspace{5mm}100 \text{\AA}$^{3}$\\[1.5mm]
& volume of a segment (C$_{8}$H$_{8}$) & $V_{2}$ & \hspace{5mm}165 \text{\AA}$^{3}$\\[1.5mm]
& Flory characteristic ratio & C$_{F}$ & \hspace{5mm}10 \,\,\,\,\,\,\,\\[1.5mm]
& mean bond length & $\bar{\ell}$ & \hspace{5mm}1.55 \text{\AA}\,\,\,\\[1.5mm]
& enthalpy parameter (25$^{\,\circ}$C) & $\chi$ & \hspace{5mm}0.4 \,\,\,\,\,\,\,\\[2mm]
\hline\\[-6mm]
\end{tabular}\\[6mm]
\end{center}
\end{table}

\begin{wrapfigure}[18]{r}{7.5cm}
\vspace{-6mm}
 \includegraphics[width=7.5cm]{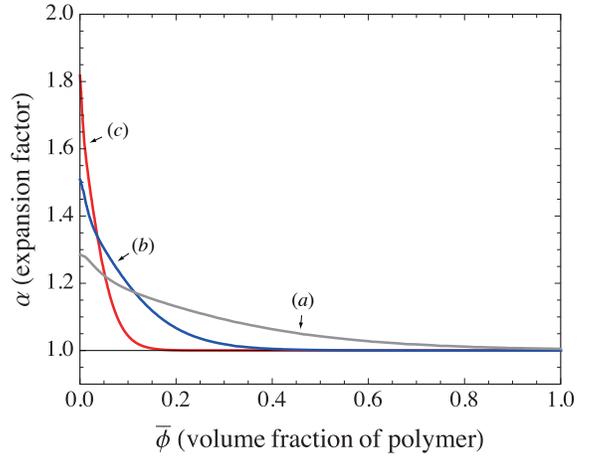}
 \vspace{-5mm}
 \caption{Molecular weight dependence of $\alpha$: (a) $\text{M}_{w}=10^{4}$, (b) $\text{M}_{w}=10^{5}$ and (c) $\text{M}_{w}=10^{6}$.}\label{fig1}
\end{wrapfigure}

The simulation results are illustrated in Fig. \ref{fig1} for M$_{w}=10^{4}$, $10^{5}$ and $10^{6}$. It is seen that the molecular weight has a marked effect on the location of the swollen-to-unperturbed coil transition point, $c^{*}, (\alpha=1$). With increasing M$_{w}$ (from (a) to (c)), $c^{*}$ shifts rapidly to lower concentration range. This phenomenon is easy to understand: Because of the form, $p(s)=(d/2\pi\langle s^{2}\rangle)^{d/2}\exp(-d\hspace{0.2mm}s^{2}/2\langle s^{2}\rangle)$, of the segment distribution around the center of gravity, chains tend to interpenetrate more deeply as M$_{w}$ increases. Thus in the limit of an infinitely long chain, the fluctuation term $J$ goes to zero, and the excluded volume effects vanish in all finite concentrations. For this reason, it is only at zero concentration that the notion of the infinite chain has sound physical basis in the study of the excluded volume effects. In contrast, for short chains, the excluded volume effects tend to survive in high concentration region (curve (a)), suggesting that for chains less than M$_{w}=10^{4}$, the excluded volume effects may not disappear even at the melt state ($\bar{\phi}=1$).

\begin{wrapfigure}[14]{r}{7.5cm}
\vspace{-8.5mm}
 \includegraphics[width=7.5cm]{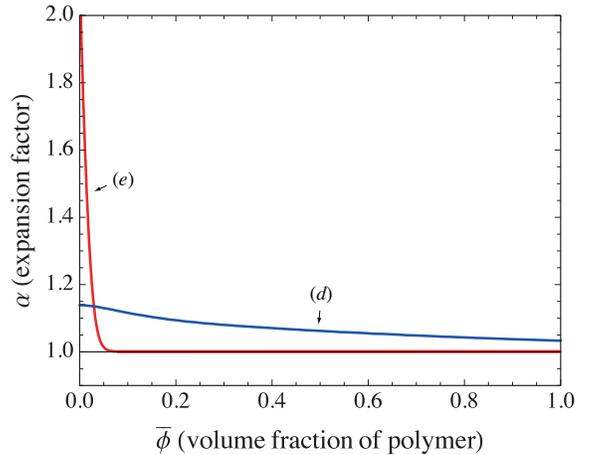}
  \vspace{-5mm}
 \caption{Molecular weight dependence of $\alpha$: (d) $\text{M}_{w}=10^{3}$ and (e) $\text{M}_{w}=10^{7}$.}\label{fig2}
\end{wrapfigure}

To examine the above inference, we have calculated two extreme cases of M$_{w}=10^{3}$ and $10^{7}$; the results are shown in Fig. \ref{fig2}. As expected, for the short chain (M$_{w}=10^{3}$), the excluded volume effects never disappear in all concentration range from the dilution limit to the melt; the coil remains swollen even at the melt state. We know of course that no such short Gaussian chain having M$_{w}=10^{3}$ exists in reality, so our discussion is purely theoretical. Notwithstanding the present results reveal that ``the ideal chain hypothesis at the melt state (ICHM)'' is not strictly true, but a practical law valid only for polymers having M$_{w}\gtrsim10^{4}$\cite{Cotton}. The swollen-to-unperturbed coil transition point should vary from $c^{*}=\infty$ to 0 as $\text{M}_{w}$ varies from 0 to $\infty$.

The above statement can be reinforced by re-examining the original Flory theory: The local free energy

\begin{equation}
\delta F_{mixing}=kT\left\{\log(1-v_{2})+\chi\hspace{0.2mm}v_{2}\right\}\delta n_{1}\label{1-4}
\end{equation}
stands for the difference in the Gibbs potential between the pure components (polymers and solvents) and the mixture. In the point of view of polymers as solutes, eq. (\ref{1-4}) represents the potential difference between the melt state and the solution. Integrating eq. (\ref{1-4}) over all volume elements in the system, then adding the elastic term, and minimizing the resultant total free energy, we are led to the known result, $\alpha^{5}-\alpha^{3}=C\,(1-\Theta/T)\,M^{1/2}$. It becomes clear that the classic theory identifies the melt state (pure polymer) with the standard state and $\alpha$ is calculated as the difference from the melt state. It is realized that the classic theory postulates, as the premise, the ideal chain behavior at the melt state. This will be the reason, despite the fact that ICHM has been confirmed firmly by SANS experiments\cite{Cotton}, why ICHM has raised so many questions and debates so far\cite{deGennes, deCloizeaux, Wittmer}. 

Finally we would like to emphasize that, aside from the problem of the standard state, the classic theory\cite{Flory} has extracted correct and essential features of the excluded volume effects, i.e., it has made the complete description of the limiting case of $C=0$ for the generalized expression, eq. (\ref{1-1}).


\end{document}